\documentclass[preprint,proceedings]{rmaa}

\suppressfulladdresses % by popular test

\usepackage{paralist}

\usepackage{psfrag,color}

\SetYear{2007}
\SetConfTitle{The Nuclear Region, Host Galaxy and Environment
 of Active Galaxies}

\title{The big blue bump and soft X-ray excess of indivudual quasars.} 

\author{
  Sinhu\'e A.R. Haro-Corzo,\altaffilmark{1} 
  Luc Binette,\altaffilmark{2}
  and Yair Krongold\altaffilmark{2}}

\altaffiltext{1}{Instituto de Ciencias Nucleares,
  Universidad Nacional Autonoma de M\'exico, Apartado Postal 70-543,
C.P 04510,
  Ciudad de M\'exico, Estados Unidos Mexicanos.}

\altaffiltext{2}{Instituto de Astronom\'ia, Universidad Nacional Autonoma de M\'exico, Apartado Postal 70-264,
C.P 04510,
  Ciudad de M\'exico, Estados Unidos Mexicanos.}

\shortauthor{Haro-Corzo et al.}
\shorttitle{RevMexAA(SC) Demo Document}

\listofauthors{Sinhu\'e A.R. Haro-Corzo, Luc Binette \&Yair Krongold }
\indexauthor{Haro-Corzo, S.A.R.}
\indexauthor{Binette, L.}
\indexauthor{Krongold, Y.}

\abstract{
For 11 quasar, we find that the soft X-ray excess
  component is not prolongation of the Big Blue Bump. Furthermore,
  adopting a theoretical continuum that is absorbed by the appropriate
  amount of intrinsic dust, we are able to reconcile this universal
  theoretical continuum with the {\it UV break}  and the {\it softness problem}.}

\resumen{
Para 11 quasares, encontramos que 
el exceso de rayos-X suaves {\it no} es una prolongaci\'on de la Gran
Joroba Azul. A\'un m\'as, adoptando un continuo ionizante te\'orico
absorbido por una cantidad diversa de polvo intr\'inseco
para cada quasar, podemos reconciliar este continuo te\'orico con el
quiebre UV con el problema de suavidad.}

\addkeyword{dust,extinction}
\addkeyword{galaxies: ISM}
\addkeyword{quasars: general}
\addkeyword{ultraviolet: galaxies}
\addkeyword{X-ray: galaxies}

\begin{document}
% Typeset article header
\maketitle

\section{Introduction}
\label{sec:intro}
The optical-UV spectral feature in quasars commonly referred as the
{\it Big Blue Bump} (BBB) is generally believed to be the thermal
manifestation of an accretion disk.  However, state-of-art models do
not reproduce well the {\it UV break} observed near 1100\,\AA. In
addition, models that roughly reproduce the shape of the observed BBB
present the co-called {\it softness problem} as pointed out by Binette
et\,al (B07, this meeting). No disk model can solve both problems
simultaneously. In the X-ray domain, the spectra of quasars are
frequently fit with a simple power law. However, there are quasars
that show additional components such as intrinsic absorption or a soft
excess \citep{mathur94}. An observational gap exists between the EUV
and the X-rays (the gray zone in Fig.\,1 of B07) where the Milky Way
absorbs all photons. Thus the SED behavior within this gap is not
known.  The usual way of connecting the ionizing continua
of these two domains is to use a simple extrapolation of the
far-UV power law into the soft X-rays. In this work, we explore the
following questions:
\begin{inparaenum}[(1)]
\item Is the  the Soft X-ray Excess a simple prolongation of the BBB?
\item Is it possible to reconcile the {\it UV break} with a theoretical
universal SED? 
and finally,
\item Does this new intrinsic SED resolve the {\em softness problem}? 
\end{inparaenum}

\section{Matching the UV and X-ray Components}
\label{sec:sample}
In order to resolve these questions and to derive useful
constraints on  BBB models, we analyzed 
the individual UV and X-ray spectra of 11 quasars for which
high-quality spectral data sets exist in both the {\it HST-FOS}
\citep[kindly lent us by Telfer;][]{telfer02} and {\it
Chandra-ACIS} (obtained from the archives). Our figure labels are related to the quasars
names as follow:
\begin{inparaenum}[(a)]
\item PKS\,1127-14,
\item PKS\,0405-123,
\item 3C\,351,
\item 3C\,334
\item B2\,0827+24,
\item PKS\,1354+19,
\item 3C\,454.3,
\item OI\,363,
\item PKS\,1136-13,
\item PG\,1634+706,
\item PG\,1115+080.
\end{inparaenum}
These eleven quasars have the following characteristics 
\citep[see][hereafter H07]{haro07}: a redshift between 0.3
and 1.8, a sufficiently wide spectral coverage shortward and longward
of 1100 \AA\ and an an absence of any deep absorption trough. We fit
each observed domain (UV and X-rays) separately. The best-fits in the
UV are based on two different hypotheses concerning the environment of
each quasar (see below). Finally, we combine the best-fits from both
domains into a single SED in $\nu F\nu\ $ ($\propto \nu^{\beta}$,
where ${\beta=+\alpha +1}$) and attempt to extrapolate the
fits towards the EUV gap, as shown in Fig.\,\ref{fig:sed11}.

\subsection{The UV SED in the Dust-free Case}
\label{sec:dust-free}

\begin{figure*}[!t] 
\includegraphics[width=18cm,height=18cm]{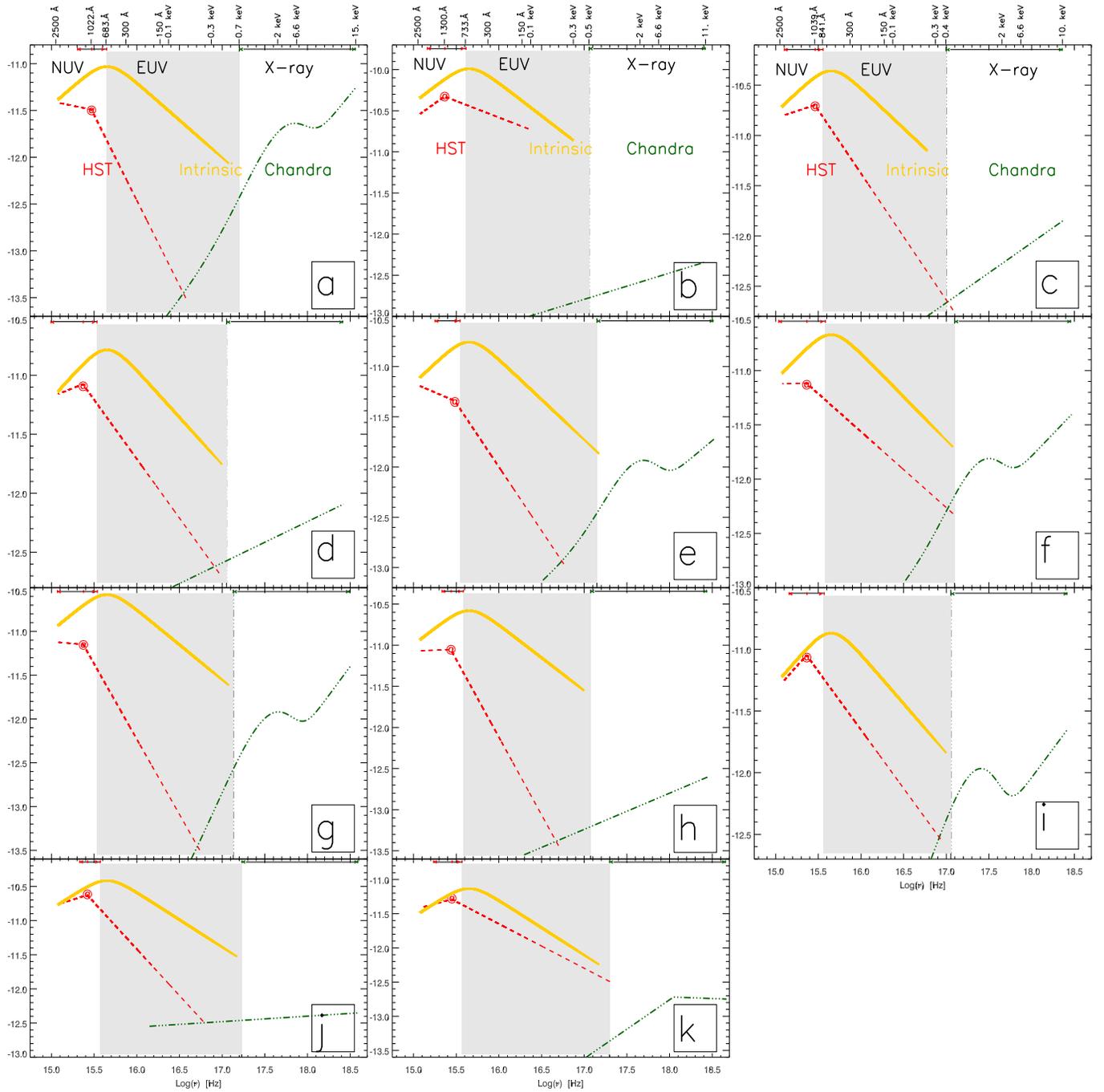} 
\caption{Best-fit models in $\nu F\nu\,vs\,\nu$ of the UV and X-ray
domains for each quasar. To avoid cluttering, the data are not superimposed
(but see H07). In the UV domain, the models are based on two
different hypotheses concerning the environment of individual quasars:
the dust-free case (dashed lines, see \S\,\ref{sec:dust-free}) or the
dust-absorbed case, by amorphous carbon and nanodiamond grains
(thick solid lines, see \S\,\ref{sec:dust}).  The X-ray continuum for
each quasar is represented by a dot-dashed line. The symbol ``@''
denotes the {\it UV break} position. The gray shaded area represents
the domain where there is no data.  Black arrows at the top of each
panel indicate the wavelength coverage of the original data.[See the
electronic edition of the Journal for a color version of this figure.]
} \label{fig:sed11}
\end{figure*}

We assume in this case that the environment is dust-free, i.e., the
observed UV spectra only need to be corrected for Galactic dust and
intergalactic Ly$\alpha$ absorption. The UV continuum best-fits using
broken powerlaws as templates are shown in Fig.~\ref{fig:sed11}
(dashed lines). The UV spectra are all characterized by a sharp {\it
  UV break} between 988\,\AA\ and 1303\,\AA, the position of which is
denoted by the symbol ``@''. The steepening in the SED shortward of
the {\it UV break} varies from object to object. In the case of the
X-ray domain, the best-fits (dot-dashed line) were corrected for
Galactic absorption. We obtained a marked improvement in the
statistics of the X-ray models, however, after considering an {\it
  intrinsic cold\footnote{Except for 3C\,351, where an ionized
    absorber is required.}} absorption component\footnote{The presence
  of intrinsic absorption allows some room for the possible presence
  of extinction in the UV, as explored below in \S~\ref{sec:dust}.} in
7 quasars (for the remaining 4 quasars, only an upper limit was
derived) and after adding a soft X-ray excess component (in 5
quasars). As can be appreciated in Figure~\ref{fig:sed11}, for 9 out
of 11 quasars, the far-UV best-fit powerlaw either does not connect
with the X-ray SED or result in an inflection in the SED. We conclude
that the soft X-ray component is not a simple prolongation of the BBB
component, but must have a distinct origin.

\subsection{The UV SED in the Dust Case}
\label{sec:dust}
The paradigm behind this section is that a universal intrinsic SED
exists in most AGN, but it is absorbed by different amounts of
intrinsic dust.  As before, Galactic and extragalactic absorptions
were considered in our data analysis. The working hypothesis behind
the proposed test are the following:

\begin{asparaitem}
\item {\bf Intrinsic Universal SED}: We have been exploring intrinsic
  SEDs (see H07) that they can produce a much higher number of hard
  ionizing photons (in the extreme-UV) than provided by the broken
  powerlaw adopted in the dust-free case
  (see\,\S\,\ref{sec:dust-free}). This new intrinsic SED (solid lines
  in Figure~\ref{fig:sed11}) could in principle help resolve the {\it
    softness problem} because it resembles the very hard SEDs used in
  theoretical studies of the BLR emission lines (plotted in Figs~8 and
  9 of H07). We define our universal SED as a power law with a near-UV
  spectral index in $\nu F\nu\ $ of $\beta_{NUV}=0.8$, multiplied by a
  rollover function defined in H07, in order that our SED remains
  consistent with the observed X-ray flux. This means that our
  spectral index in the {\it extreme} UV becomes $\beta_{FUV}=-0.8$.

\item {\bf Intrinsic Dust Screen}: Based on previous works (H07, see
also Fig.\,4 of B07), successful dust models rely on a grain
composition made of two types: nanodiamonds and amorphous
carbon. Hence two extinction curves are needed. While both are
responsible of absorbing a significant amount of UV photons, only
nanodiamonds can account for the sharp UV break.
\end{asparaitem}

We can successfully reproduce the observed spectra (F$_{obs}$) by
attenuating our universal SED (F$_{int}$) by two transmissions curves,
i.e., F$_{obs}$=F$_{int}\times exp(-N_H\,\sigma_{ND}) \times
exp(-N_H\,\sigma_{AC})$, where $N_H$ is a free parameter, that
controls the amount of dust required to reproduce the spectral index
observed. Interestingly, we have found that the column densities of
the dust screen are comparable to those inferred from the X-ray
best-fits (see H07).  For all quasars, we find that our extrapolation
of the intrinsic SED lies above the observed X-ray flux. This is shown
in Fig.\,\ref{fig:sed11}. This result implies that the ``true''
universal SED should be characterized by a sharper turn-over or one
that occurs at much higher energies than assumed here. Maybe accretion
disk models might be more suitable than our assumed ad\,hoc universal
SED given that they are meanfully and likely to provide the required
photons and luminosity expected by the photoionization models.

\section{Discussion}
We found that it is possible to account for the presence of a sharp UV
break and to resolve the {\it softness problem} by having just a
universal intrinsic SED attenuated by dust. This universal SED can
also be made compatible with those required by photoionization
calculations of the BLR.

For almost all of the 11 quasars (with or without dust absorption),
the soft X-ray component must be different from the BBB, given that an
extrapolation of the BBB powerlaw results in a sharp SED inflection in
the EUV at the junction point with the X-ray best-fit models.

\acknowledgements SARHC acknowledges current support from a CONACyT
postdoctoral fellowship and PAPIIT grant 108207. LB and YK are funded
by the CONACyT grant J-49594, J-50296 and Papiit 118905. We
acknowledge the technical support of Alfredo D\'{\i}az Azuara and
Carmelo Guzman.

\bibliography{sed}

% \begin{thebibliography}
% \bibitem{}Abbott, D. C., Bieging, J. H., Churchwell, E., \& Cassinelli, J. P.
%  1980, ApJ, 238, 196
% \end{thebibliography}

\end{document}